\documentclass[final,1p,times]{elsarticle} 

\usepackage{graphicx}
\usepackage{amssymb} 
\usepackage{amsthm} 
\usepackage{lineno}

\journal{Nuclear Physics A} 

\begin{document}

\begin{frontmatter} 

\title{The QCD equation of state with 2+1 flavors of\\
Highly Improved Staggered Quarks (HISQ)}

\author{Alexei Bazavov (for the HotQCD\fnref{col1} 
Collaboration)}
\fntext[col1] {The HotQCD collaboration members are:
A.~Bazavov,
T.~Bhattacharya,
M.~Buchoff,
M.~Cheng,
N.~Christ,
C.~DeTar,
H.-T.~Ding,
S.~Gottlieb,
R.~Gupta,
P.~Hegde,
U.~Heller,
C.~Jung,
F.~Karsch,
E.~Laermann,
L.~Levkova,
Z.~Lin,
R.~Mawhinney,
S.~Mukherjee,
P.~Petreczky,
D.~Renfrew,
C.~Schmidt,
C.~Schroeder,
W.~Soeldner,
R.~Soltz,
R.~Sugar,
D.~Toussaint,
P.~Vranas.}
\address{Brookhaven National Laboratory, Upton, NY 11793}


\begin{abstract}
One of the fundamental properties of the 
quark-gluon plasma (QGP), the 
equation of state, is a subject of extensive studies in lattice QCD
and an essential requirement for the correct
hydrodynamic modeling of heavy-ion collisions. Lattice QCD provides 
first-principle calculations for the physics in the non-perturbative
regime. In this contribution,
we report on recent progress by the HotQCD collaboration in 
studying the 2+1 flavor equation of state on lattices with the temporal 
extent $N_\tau=6$, $8$, $10$ and $12$ in Highly Improved Staggered 
Quarks~(HISQ) discretization scheme. Comparisons with 
equation of state calculations with different fermion actions are also
discussed.
\end{abstract}

\end{frontmatter} 


At fixed cutoff (in our case, lattice spacing, $a$) the trace of the 
energy-momentum tensor, or interaction measure, can be related
to the partition function of the system:
\begin{equation}
\varepsilon-3p = -\frac{T}{V}\frac{d\ln Z}{d\ln a},\,\,\,\,\,
Z=\int DU D\bar\psi D\psi \exp\left(-S_{gauge}-S_{fermion}\right),
\end{equation}
where $\varepsilon$ is the energy density and $p$ is pressure.

The equation of state, \textit{i.e.}, the dependence of pressure
on temperature can be expressed then as an integral:
\begin{equation}
    \frac{p}{T^4}-\frac{p_0}{T_0^4}=\int_{T_0}^T
    dT'\frac{\varepsilon - 3p}{T'^5}.
\end{equation}

To calculate the interaction measure in 2+1 flavor QCD we work
on a Euclidean space-time lattice and evaluate path integrals with
the importance sampling technique. We use a tree-level improved
action for gauge fields and the Highly Improved 
Staggered Quarks (HISQ) action for fermions \cite{HPQCD}. To get
finite values, subtraction of UV divergences is required:
\begin{equation}\label{tram}
  \frac{\varepsilon-3p}{T^4} = R_\beta[\langle S_{gauge}\rangle_0-
  \langle S_{gauge}\rangle_T]
  -R_\beta R_m[2m_l(\langle \bar l l\rangle_0-
  \langle \bar l l\rangle_T)+m_s(\langle \bar s s \rangle_0-
  \langle \bar s s\rangle_T)],
\end{equation}
\begin{figure}[htbp]
\begin{center}
\includegraphics[width=0.495\textwidth]{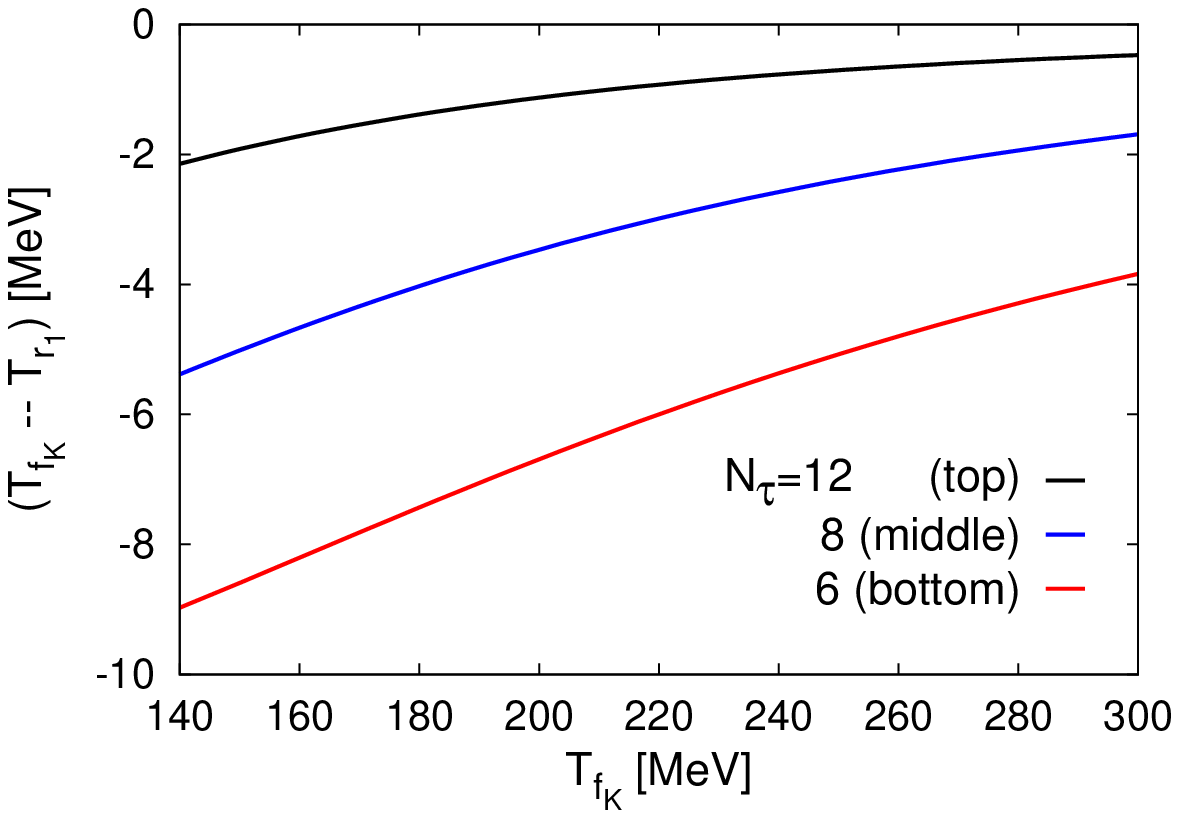}\hfill
\includegraphics[width=0.462\textwidth]{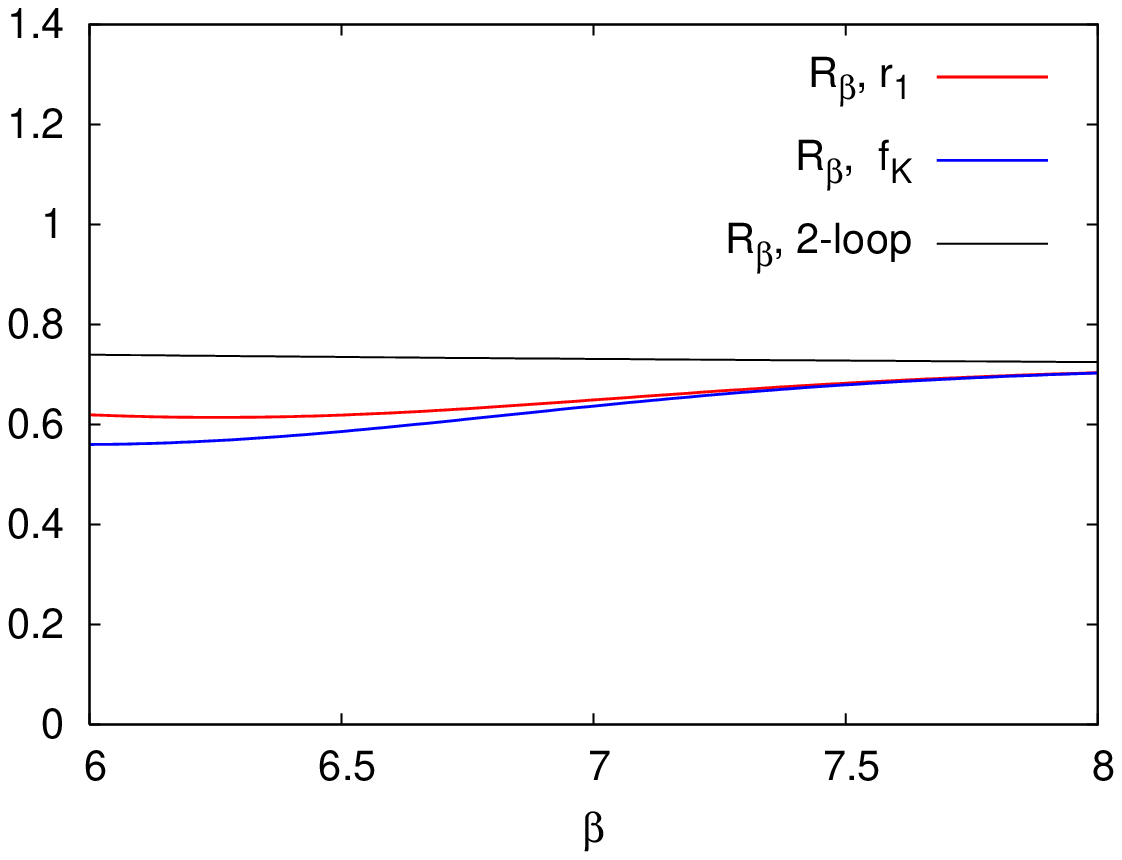}
\end{center}
\vspace{-7mm}
\caption{The difference in temperature (left) set by $r_1$ and $f_K$
scale, described in the text. The running of the gauge coupling (right).}
\vspace{-3mm}
\label{fig_Tr1}
\end{figure}
where the subscript ``$0$'' refers to $T=0$ and ``$T$'' to finite
temperature for observables evaluated at the same value of the cutoff.
On a hyper-cubic lattice $N_s^3\times N_\tau$ the physical temperature
is set by the size of the temporal dimension and the lattice spacing
as $T=1/(N_\tau a)$. For $T=0$ calculations we use $N_\tau \geqslant N_s$,
and for $T>0$ we keep $N_s/N_\tau=4$ and at fixed $N_\tau$ vary the
lattice spacing $a$ by varying the gauge coupling $\beta=10/g^2$.
The continuum limit in this setup is controlled by $N_\tau\to\infty$,
therefore, we carried out this study on several sets of ensembles with
$N_\tau=6$, $8$, $10$ and $12$. The strange quark mass $m_s$ is tuned
to the physical value, while the two degenerate light quarks have masses
$m_l=m_s/20$, slightly heavier than physical ($m_l\simeq m_s/27$).
To set the lattice spacing in physical units (fm) we use the Sommer-type
scale \cite{Sommer_scale} $r_1=0.3106$~fm \cite{MILC_r1},
or, alternatively, the kaon decay constant, $f_K=156.1$~MeV.
From Fig.~\ref{fig_Tr1} (left) one can
see what effect using different reference observables has on converting
the temperature from lattice units to MeV. Over the temperature range of
interest on $N_\tau=6$ lattices the difference is within $9$~MeV, and
on $N_\tau=12$ within $2$~MeV. Another effect of using different
scales is related to running of the corresponding $\beta$-functions:
\begin{equation}
R_\beta(\beta)=-a\frac{d\beta}{da},\,\,\,\,\,
R_m(\beta)=\frac{1}{m_s(\beta)}\frac{dm_s(\beta)}{d\beta},
\end{equation}
where $m_s(\beta)$ defines a line of constant physics (LCP), \textit{i.e.},
such combination of the gauge coupling and the strange quark mass so that
pion and kaon masses (in MeV) stay approximately
constant in the whole $\beta$ range used in the simulation.
\begin{figure}[htbp]
\begin{center}
\includegraphics[width=0.495\textwidth]{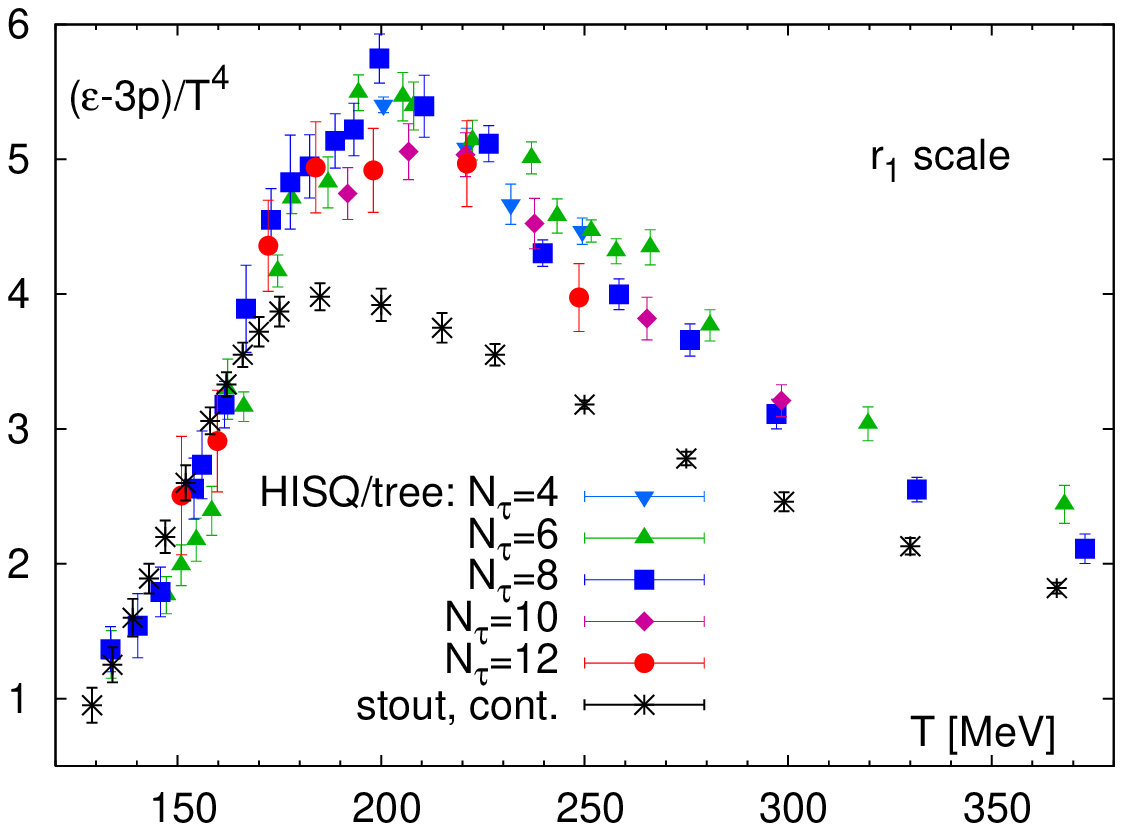}\hfill
\includegraphics[width=0.495\textwidth]{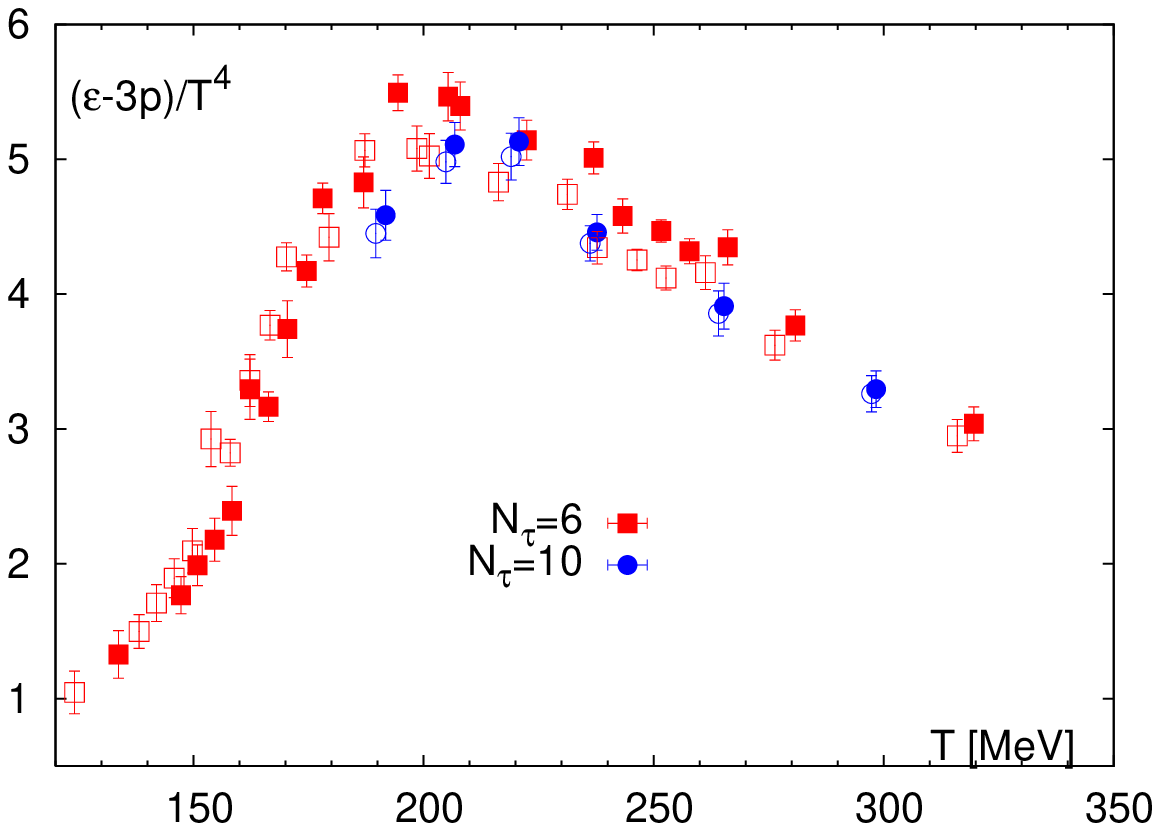}
\end{center}
\vspace{-7mm}
\caption{Comparison of the HISQ/tree interaction measure on 
$N_\tau=6$, $8$, $10$ and $12$ lattices with the stout continuum
estimate~\cite{stout_eos} (left), $N_\tau=6$ and $10$ HISQ/tree
data (right). Filled (open) symbols in the right panel correspond 
to the $r_1$ ($f_K$) scale, see text. }
\vspace{-3mm}
\label{fig_EoS_hisq}
\end{figure}
$R_\beta$ for $r_1$ and $f_K$ scale is shown in Fig.~\ref{fig_Tr1} (right),
together with the 2-loop perturbative result. As is clear from Eq.~(\ref{tram}),
the $\beta$-functions enter into the interaction measure multiplicatively.

Present status of the interaction measure with HISQ/tree is shown
in Fig.~\ref{fig_EoS_hisq} together with the stout continuum estimate of
Ref~\cite{stout_eos}. Although we have not yet performed
the continuum limit, the HISQ/tree results seem to disagree with stout
in $T=170-350$~MeV range.

\begin{figure}[htbp]
\begin{center}
\includegraphics[width=0.495\textwidth]{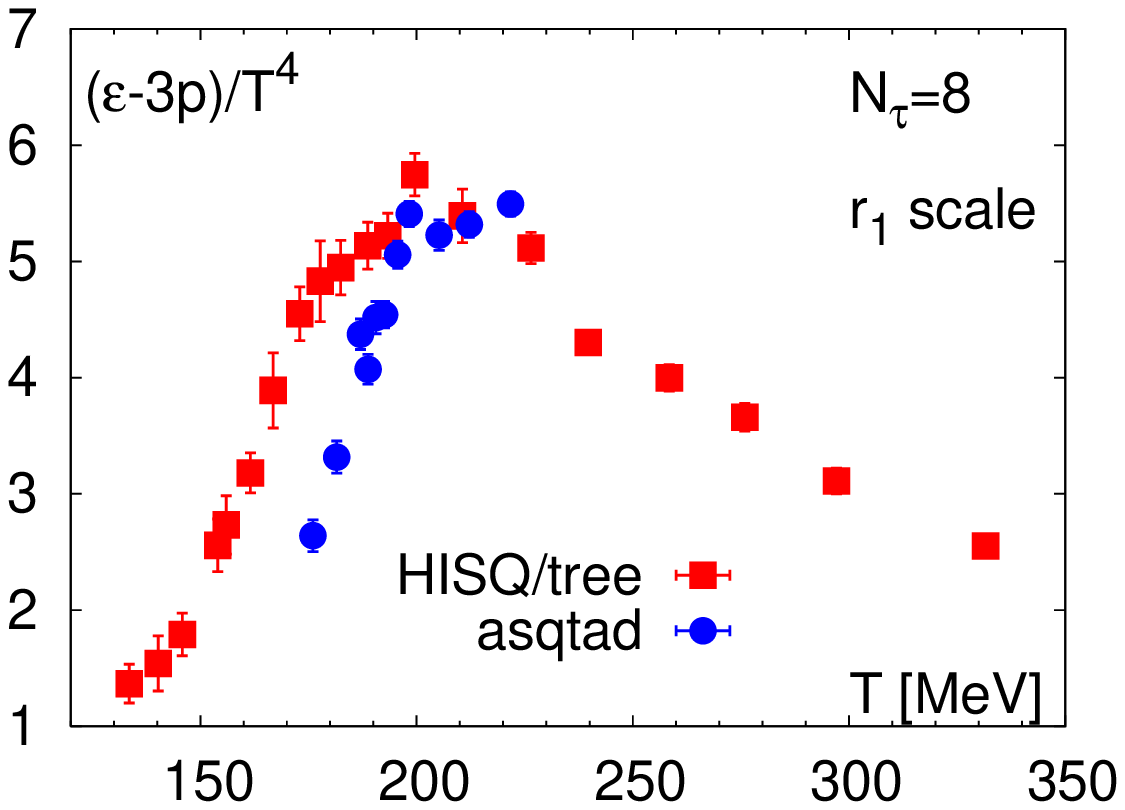}\hfill
\includegraphics[width=0.495\textwidth]{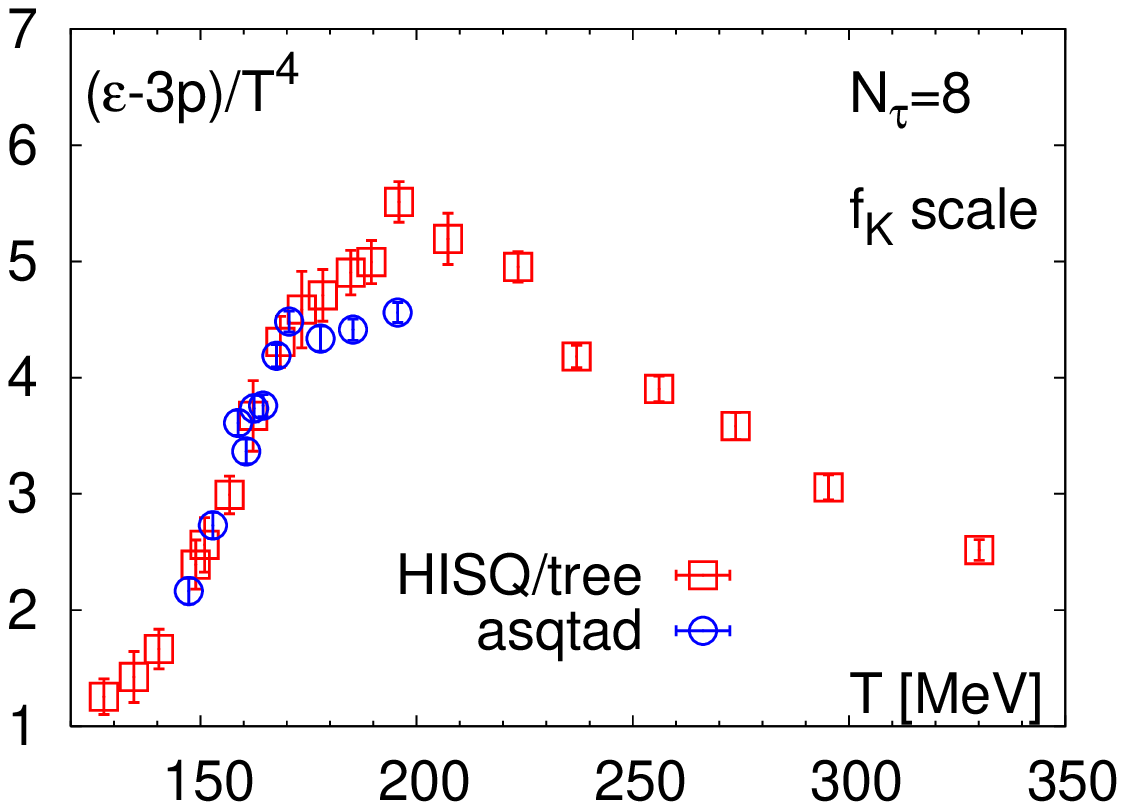}
\includegraphics[width=0.495\textwidth]{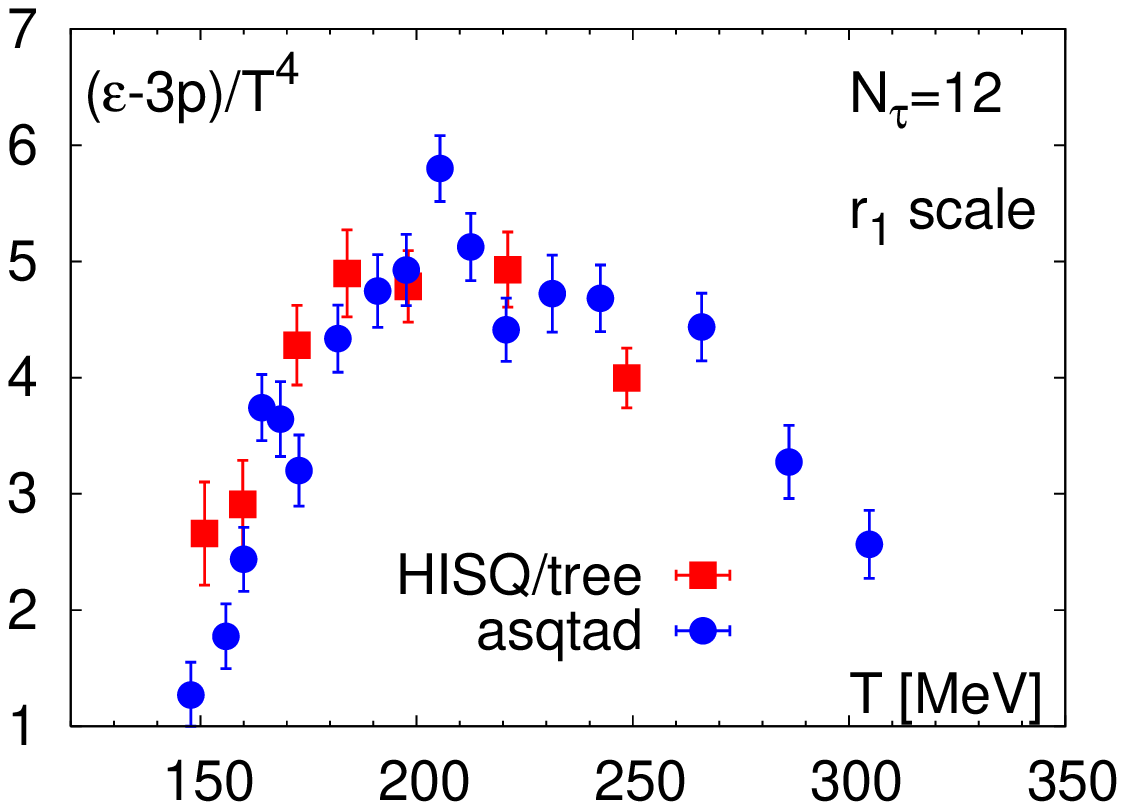}\hfill
\includegraphics[width=0.495\textwidth]{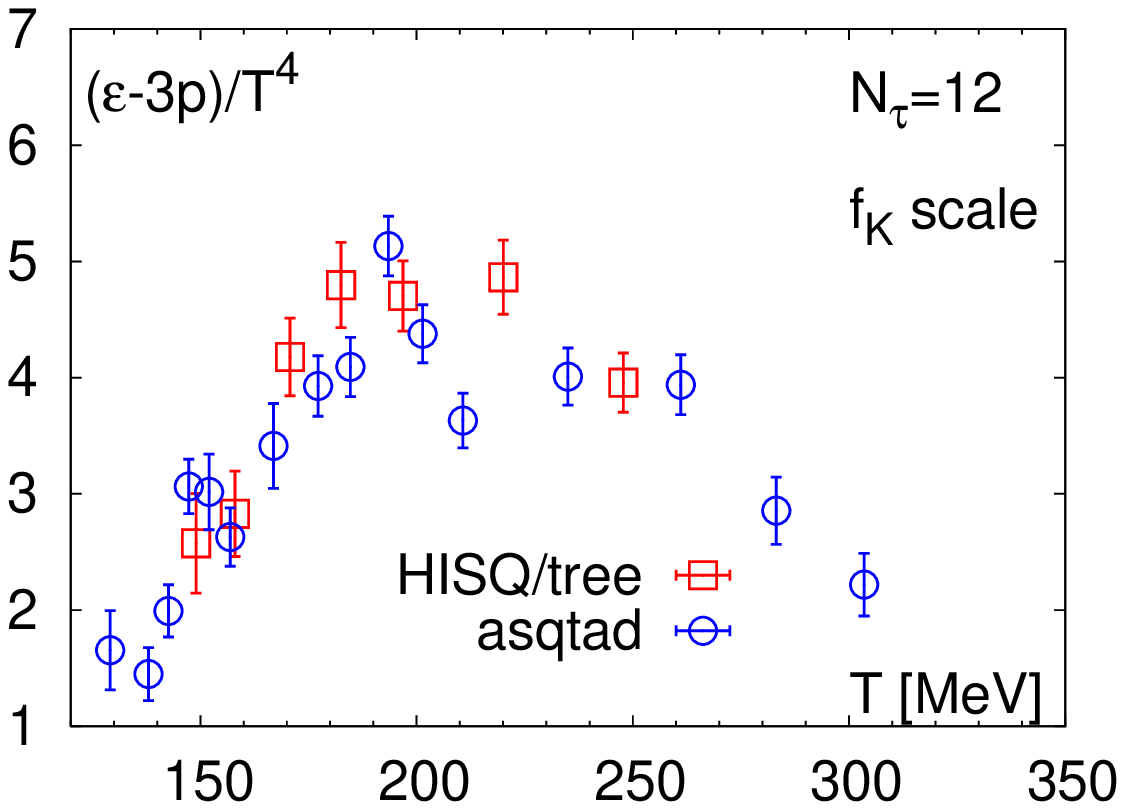}
\end{center}
\vspace{-7mm}
\caption{The interaction measure for the HISQ/tree action on
$N_\tau=8$ (top) and $N_\tau=12$ (bottom) ensembles with $r_1$ (left)
and $f_K$ (right) scale.}
\vspace{-3mm}
\label{fig_hisq_12}
\end{figure}

The interaction measure for $N_\tau=6$ and $10$
is shown in Fig.~\ref{fig_EoS_hisq} (right),
$N_\tau=8$ and $N_\tau=12$ in Fig.~\ref{fig_hisq_12}.
We also include our data with the asqtad action, where available.
For comparison we plot
the HISQ/tree and asqtad results with the temperature scale set with
both $r_1$ (filled symbols) and $f_K$ (open symbols).

On the coarsest lattice, $N_\tau=6$, Fig.~\ref{fig_EoS_hisq} (right)
the cutoff effects for the HISQ/tree action
are easily observed: the data plotted with
$f_K$ scale (red open symbols) lies lower and is shifted to the
left, compared to the data with the $r_1$ scale (red filled symbols).
On $N_\tau=8$ the cutoff effects for HISQ/tree are milder, but
still are very substantial for the asqtad action, compare the filled
blue symbols in the left panel to the open blue symbols in the
right panel of the upper part of Fig.~\ref{fig_hisq_12}. On $N_\tau=10$ and $12$
for both HISQ/tree and asqtad, Fig.~\ref{fig_EoS_hisq} (right)
and the lower part of Fig.~\ref{fig_hisq_12}, the cutoff effects are 
comparable with the statistical 
errors and agreement between the two actions is good for
$r_1$ and $f_K$ scale.

\begin{figure}[htbp]
\begin{center}
\includegraphics[width=0.495\textwidth]{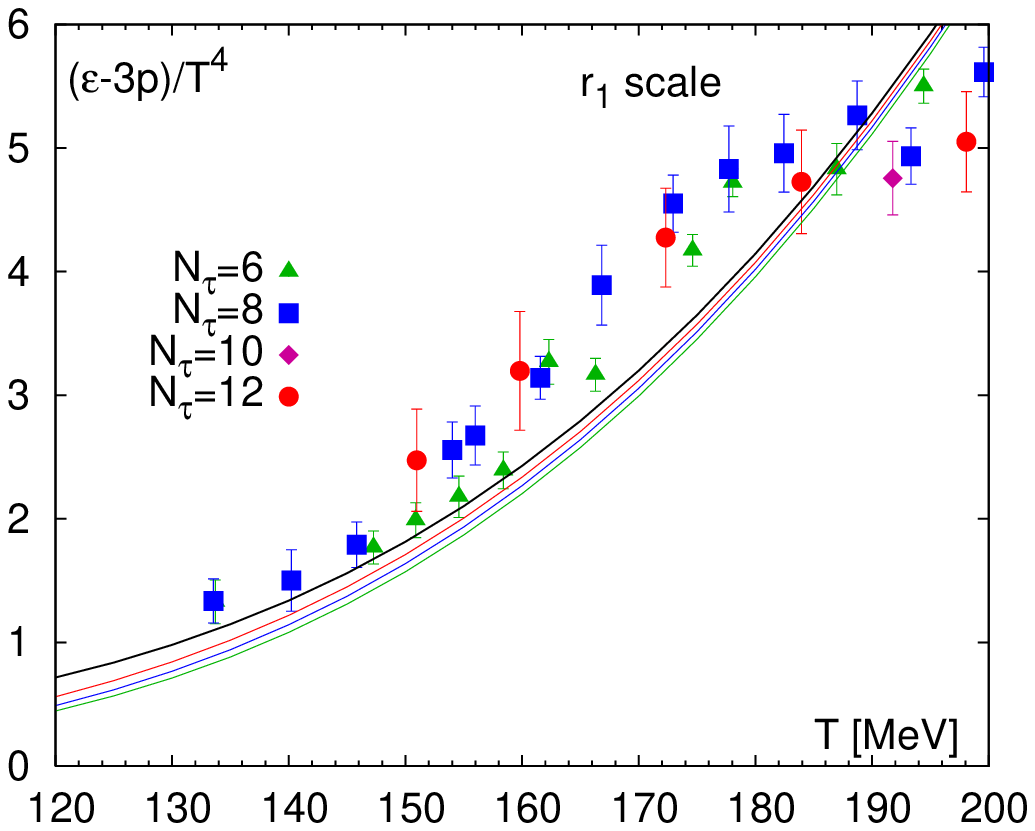}\hfill
\includegraphics[width=0.495\textwidth]{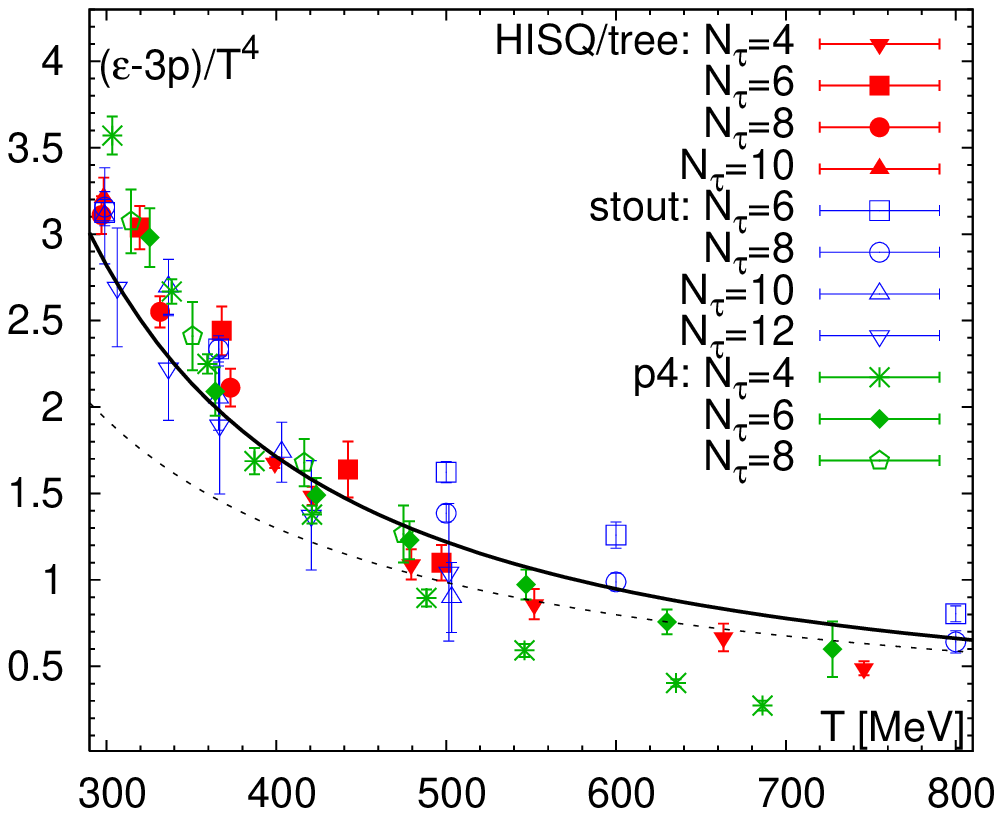}
\end{center}
\vspace{-7mm}
\caption{The interaction measure at low (left) and high (right)
temperature. The curves on the left panel represent the HRG calculation,
see text. On the right the dashed curve shows 1-loop and the solid
curve 2-loop perturbative results.}
\vspace{-3mm}
\label{fig_hisq_high}
\end{figure}

The low-temperature behavior of the interaction measure is shown in
Fig.~\ref{fig_hisq_high} (left). The lines represent the hadron
resonance gas (HRG) calculation with the physical pion and
also with the full unphysical heavier pion multiplet, as encountered
on the lattice. (The effects of heavier pion multiplet at fixed
lattice spacing for the HISQ/tree action are quantitatively
addressed in Ref.~\cite{hisq_lat10}.)
There are two interesting observations:
1. Although the root-mean-squared pion mass varies from $400$ to
$200$~MeV for $N_\tau=6-12$ ensembles, the interaction measure in HRG
is not very sensitive to this variation. Thus, better agreement
of the HISQ/tree lattice data for various $N_\tau$ in the low-temperature
region is due to both, improved properties
of the HISQ/tree action and little sensitivity of this observable
to the cutoff effects in the pion sector. 2. The lattice result
starts to disagree with the HRG model at $T\sim150-160$~MeV.

The interaction measure at high temperature is presented in
Fig.~\ref{fig_hisq_high} (right) for the HISQ/tree, p4 and stout
action. The lines represent perturbative calculations.
For $T$ up to $800$~MeV cutoff effects expected in the $T\to\infty$
limit are observed only for the p4 action.

\section{Conclusion}

We continued the calculation of the 2+1 flavor QCD equation of state with
the HISQ/tree action on $N_\tau=6$, $8$, $10$ and $12$ ensembles.
Using different observables to set the scale, $r_1$ and $f_K$, allows
for a crude estimate of the magnitude of cutoff effects, which seem
to be small at the finest $N_\tau=12$ lattices. Better control over
the statistics on $N_\tau=10$ and $12$ ensembles is required before
an extrapolation to the continuum may be attempted.

\section*{Acknowledgments}

The numerical simulations
have been performed on BlueGene/L computers at Lawrence Livermore
National Laboratory (LLNL), the New York Center for Computational
Sciences (NYCCS) at Brookhaven National Laboratory, 
US Teragrid (Texas Advanced Computing
Center), Cray XE6 at the National Energy Research Scientific
Computing Center (NERSC),
and on clusters of the USQCD collaboration in JLab and FNAL.

\end{document}